\documentclass[twocolumn,aps,prc,showpacs,superscriptaddress,preprintnumbers,amssymb,floatfix,nofootinbib]{revtex4}
\usepackage{epsfig,graphics}
\usepackage{graphicx}
\usepackage{dcolumn}
\usepackage{bm}
\usepackage{amsmath}
\usepackage[usenames]{color}
\usepackage{ulem} 
\usepackage[colorlinks,linkcolor=blue,urlcolor=blue,citecolor=blue]{hyperref}
\usepackage{hyperref}
\usepackage{tikz}
\usetikzlibrary {shapes.geometric}
\usetikzlibrary{shapes,arrows}
\begin{document}

\title{Machine learning study to identify collective flow in small and large colliding systems}
\author{Shuang Guo}
\affiliation{Key Laboratory of Nuclear Physics and Ion-beam Application (MOE), Institute of Modern Physics, Fudan University, Shanghai 200433, China}
\affiliation{Shanghai Research Center for Theoretical Nuclear Physics, NSFC and Fudan University, Shanghai $200438$, China}
\author{Han-Sheng Wang}
\affiliation{Key Laboratory of Nuclear Physics and Ion-beam Application (MOE), Institute of Modern Physics, Fudan University, Shanghai 200433, China}
\affiliation{Shanghai Research Center for Theoretical Nuclear Physics, NSFC and Fudan University, Shanghai $200438$, China}
\author{Kai Zhou}
\email[]{zhoukai@cuhk.edu.cn}
\affiliation{School of Science and Engineering, The Chinese University of Hong Kong, Shenzhen (CUHK-Shenzhen), Guangdong, $518172$, China}
\author{Guo-Liang Ma}
\email[]{glma@fudan.edu.cn}
\affiliation{Key Laboratory of Nuclear Physics and Ion-beam Application (MOE), Institute of Modern Physics, Fudan University, Shanghai 200433, China}
\affiliation{Shanghai Research Center for Theoretical Nuclear Physics, NSFC and Fudan University, Shanghai $200438$, China}


\begin{abstract}

Collective flow has been found to be similar between small colliding systems ($p$ $+$ $p$ and $p$ $+$ A collisions) and large colliding systems (peripheral A $+$ A collisions) at the CERN Large Hadron Collider. In order to study the differences of collective flow between small and large colliding systems, we employ a point cloud network to identify $p$ $+$ Pb collisions and peripheral Pb $+$ Pb collisions at $\sqrt{s_{NN}} =$ 5.02 TeV generated from a multiphase transport model (AMPT). After removing the discrepancies in the pseudorapidity distribution and the $p_{\rm T}$ spectra, we capture the discrepancy in collective flow. Although the verification accuracy of our PCN is limited due to similar event-by-event distributions of elliptic and triangular flow, we demonstrate that collective flow between $p$ $+$ Pb collisions and peripheral Pb $+$ Pb collisions becomes more distinct with increasing final hadron multiplicity and parton scattering cross section. This study not only highlights the potential of PCN techniques in advancing the understanding of collective flow in varying colliding systems, but more importantly lays the groundwork for the future PCN-related research.

\end{abstract}


\maketitle

\section{INTRODUCTION}
\label{introduction}

With advancements in computational hardware and algorithms, machine learning (ML) techniques have become powerful tools for extracting information from big data. Among many ML architectures, the point cloud network (PCN) stands out for its efficiency and effectiveness in solving problems involving point cloud-structured data, such as three-dimensional (3D) object segmentation and scene semantic parsing~\cite{Qi:2017pn}. The unique operations employed in the PCN enhance its capability to process high-dimensional data effectively compared to traditional convolutional neural networks.

The PCN has demonstrated great potential in handling complex and irregular data in various physics domains. It excels in situations where traditional methods struggle with geometric complexity and high-dimensional data. For example, in fluid dynamics, the PCN is used to predict fluid flow fields on irregular geometries. By using point-cloud-based neural networks, researchers are able to capture geometric features and solve the partial differential equations governing fluid flow, resulting in accurate predictions in complex domains~\cite{kashefi2021point, kashefi2022physics}. In quantum computing, the integration of quantum convolutional neural networks with point cloud data processing provides a scalable solution for classifying high-dimensional data, thereby advancing quantum machine learning~\cite{baek20223d}. Additionally, in materials science, especially in additive manufacturing, the PCN can detect surface defects in real-time during the manufacturing process, leading to improved product quality and manufacturing efficiency~\cite{chen2021rapid}.

Recently, machine learning (ML) techniques have been widely used in various areas of high-energy nuclear physics~\cite{Pang:2016vdc,Zhou:2018ill,Graczykowski:2022zae,Meng:2022ssp,LingXiao:2022ssp,WanBing:2022ssp,YiLun:2022ssp,fupeng:2022ssp,He:2023urp,Ma:2023zfj,Zhou:2023pti,Boehnlein:2021eym}. In high-energy nuclear physics, both experimental detector outputs and most models for heavy-ion collisions primarily produce particle tracks. Since a track can be represented as an N-dimensional point, an event naturally becomes a point cloud of disordered points in space. Consequently, PCNs have been well-utilized in this field for tasks such as precisely reconstructing collision impact parameters~\cite{Steinheimer:2019iso, OmanaKuttan:2020brq,OmanaKuttan:2021axp}, classifying equations of state (EOS)~\cite{OmanaKuttan:2020btb}, and identifying weak intermittency signals associated with critical phenomena~\cite{Huang:2021iux} by learning the data of tracks in heavy-ion collisions. The potential applications of the PCN extend far beyond the examples provided above. In this paper, we demonstrate how the PCN can distinguish between $p$ $+$ Pb collisions and peripheral Pb $+$ Pb collisions, highlighting its ability to diagnose differences in the physical characteristics of these two collision systems.

Quark-gluon plasma (QGP) at extreme conditions of high temperature and density is thought to be a form of the early universe, which has been produced in the laboratory by relativistic heavy-ion collisions at the BNL Relativistic Heavy Ion Collider (RHIC) and the CERN Large Hadron Collider (LHC)~\cite{STAR:2005gfr,PHENIX:2004vcz,ALICE:2008ngc,Bzdak:2019pkr,Luo:2017faz}. The experimental results have shown that this new type of nearly perfect fluid can translate initial spatial geometry or initial energy density fluctuations into momentum anisotropy of final particles through the pressure gradient in hydrodynamics~\cite{Heinz:2013th,Song:2007ux,Jeon:2015dfa,Shen:2020mgh,Gale:2013da,Yan:2017ivm,Alver:2010gr,Ma:2010dv,Yan:2014afa,Qin:2010pf}, thus resulting in the emergence of strong collective flow in A $+$ A collisions~\cite{Ollitrault:1992bk,Stoecker:2004qu}.

Over the last decade,  measurements of collective flow in various colliding systems, such as $p$ $+$ $p$, $p$ $+$ Pb collisions at the LHC~\cite{CMS:2010ifv,CMS:2012qk,ALICE:2012eyl,ATLAS:2012cix,CMS:2017xnj,ALICE:2019zfl}, $p$ $+$ Au, $d$ $+$ Au, and $^3$He $+$ Au collisions at RHIC~\cite{PHENIX:2014fnc,STAR:2015kak,PHENIX:2018lia,PHENIX:2021ubk,PHENIX:2022nht}, have been performed. Surprisingly, similar collective flow is found in peripheral A $+$ A and high multiplicity $p$ $+$ A collisions at the same multiplicity, raising doubts about whether QGP droplets can also be generated in small colliding systems. Many theoretical efforts have been made to understand the origin of collective flow in small colliding systems~\cite{Dusling:2015gta,Loizides:2016tew,Nagle:2018nvi}. Similar to large colliding systems, hydrodynamics in the final state can transform initial geometric asymmetry into final momentum anisotropic flow through the pressure gradient of the QGP in small colliding systems~\cite{ Bozek:2011if,Bzdak:2013zma,Shuryak:2013ke,Qin:2013bha,Bozek:2013uha,Bozek:2015swa,Song:2017wtw}. Conversely, it is generally believed that the transport model will behave more like hydrodynamics as the multiplicity or scattering cross section increases, i.e., the change of dynamics from non-equilibrium to equilibrium. A multi-phase transport (AMPT) model \cite{Lin:2004en} is capable of describing the experimental data on both radial and anisotropic flow in both large \cite{Chen:2006ub,Lin:2014tya,Ma:2016fve} and small colliding systems \cite{Bzdak:2014dia,OrjuelaKoop:2015jss,Ma:2016bbw}. Since most of partons are not scattered especially for the small colliding systems at RHIC and the LHC, a parton escape mechanism has been proposed to explain the formation of azimuthal anisotropies in the transport model~\cite{He:2015hfa,Lin:2015ucn}.  However, it has been shown that parton scatterings are crucial for generating anisotropic flows~\cite{Ma:2016bbw,Ma:2021ror}. Using a new test-particle method, we recently proved that collectivity established by final state parton scatterings is much stronger in large colliding systems than that in small colliding systems~\cite{Wang:2022rdh}. The event-averaged flow $v_n$ primarily reflects the averaged hydrodynamic response to the initial collision geometry of the produced QGP. More helpful information, such as the event-by-event (EbyE) fluctuations of the overlap region~\cite{Alver:2010gr}, can be obtained by measuring event-by-event $v_n$ distribution $P(v_n)$ for charged hadrons, as measured by the ATLAS Collaboration using an unfolding method in Pb $+$ Pb collisions at $\sqrt{s_{NN}}$ = 2.76 TeV~\cite{ATLAS:2013xzf}. This provides a good constraint on the initial condition in A $+$ A collisions. However, the corresponding experimental measurements in small colliding systems have yet to be available, which would provide more information on the differences in the origin of collectivity between small and large colliding systems.

In addition to anisotropic flow $v_n$, investigating other observables' differences between large and small colliding systems, such as pseudorapidity distribution and $p_{\rm T}$ spectra, is worthwhile.
For example, previous experimental studies have shown power law-shaped $p_{\rm T}$ spectra in small colliding systems~\cite{STAR:2006axp,ALICE:2013wgn,ALICE:2014xsp,ALICE:2020nkc,STAR:2008med,ALICE:2013rdo}, unlike exponential-shaped $p_{\rm T}$ spectra in large colliding systems~\cite{PHENIX:2003wtu,STAR:2008med,ALICE:2013rdo}.

These motivate us to adopt the PCN in supervised training, to find the different EbyE features between the final state of $p$ $+$ Pb collisions and peripheral Pb $+$ Pb collisions from the multi-phase transport (AMPT) model. Our goal is to identify the discrepancies in collective flow between large and small systems, which will enhance our understanding of the mechanisms underlying the generation of collective flow in small systems.Ultimately, we aim to combine experimental data to determine whether small colliding systems also produce the QGP or exhibit very different characteristics of collective flow compared to large colliding systems.

Additionally, our PCN study holds potential as a valuable tool for various other research interests, such as how to potentially classify the origin of atmospheric particle showers in cosmic ray physics~\cite{Erdmann:2017str,Erdmann:2019nie}. When a particle of cosmic radiation, such as a proton or a nucleus, interacts with a molecule's nucleus in the atmosphere, it generates numerous secondary particles, resulting in an air shower. These interactions are likely to be either $p$ $+$ Pb collision type or Pb $+$ Pb collision type~\cite{Lingenfelter:2019xxn}. This is the current focus of the experiments in the Telescope Array Project~\cite{Tokuno:2012mi} and the Pierre Auger Observatory~\cite{PierreAuger:2004naf}. Our PCN study could be used in cosmic ray physics to enable a better understanding of the origin of these showers.

The paper is organized as follows. First, we introduce the AMPT model, which generates the data for $p$ $+$ Pb collisions and peripheral Pb $+$ Pb collisions, and the relevant observables in Sec.~\ref{model}. In Sec.~\ref{PCN}, we describe the details of the PCN. In Sec.~\ref{results}, the test accuracy of the PCN is presented, and the relevant physics is discussed. Finally, we summarize and give the implications of our results in Sec.~\ref{summary}.

\section{MODEL AND METHOD}
\label{model}
 \subsection{A multiphase transport model}
\label{AMPT}

The string melting version of the AMPT model consists of four main stages of heavy-ion collisions, i.e., initial state, parton cascade, hadronization, and hadronic rescatterings. The initial state with fluctuating initial conditions is generated by the heavy ion jet interaction generator (HIJING) model~\cite{Gyulassy:1994ew}. In HIJING model, minijet partons and excited strings are produced by hard and soft processes, respectively. In the string melting mechanism, all excited hadronic strings in the overlap volume are converted to partons according to the flavor and spin structures of their valence quarks \cite{Lin:2001zk}. The initial positions of partons originating from melted strings are determined by tracing their parent hadrons along straight-line trajectories.
The interactions among partons are described by the Zhang's parton cascade (ZPC) model \cite{Zhang:1997ej}, which includes only two-body elastic scatterings with a g+g $\rightarrow $ g+g cross section, i.e.,
\begin{equation}\label{sigma}
\frac{d\sigma }{d\hat{t}}=\frac{9\pi \alpha ^{2}_{s}}{2}(1+\frac{\mu^2 }{\hat{s}})\frac{1}{(\hat{t}-\mu^2)^2},
\end{equation}
where $\alpha_s$ is the strong coupling constant (taken as 0.33), while $\hat{s}$ and $\hat{t}$ are the usual Mandelstam variables. The effective screening mass $\mu$ is taken as a parameter in ZPC for adjusting the parton scattering cross section. 
Note that previous AMPT model studies have shown that a parton scattering cross section of 3 mb can well describe both large and small colliding systems at RHIC and the LHC energies~\cite{Lin:2014tya,OrjuelaKoop:2015jss,Ma:2016bbw,Ma:2016fve,He:2017tla,Lin:2021mdn}. However, we will employ different parton scattering cross sections of 0 mb, 3 mb and 10 mb in order to establish collective flow of different strengths in this study.  A quark coalescence model is used for hadronization at the freezeout of the parton system. The hadronic scatterings in the hadronic phase are simulated by a relativistic transport (ART) model \cite{Li:1995pra}.  

In this study, we simulated 2 million events each for peripheral Pb $+$ Pb collisions and $p$ $+$ Pb minimum bias collisions at $\sqrt{s_{NN}}$ = 5.02 TeV, using the string-melting version of the AMPT model.

\subsection{Anisotropic flow}
\label{flow}
The anisotropic collective flow can be defined according to a Fourier decomposition of the azimuthal angle $\phi$ distribution of measured particles,
\begin{eqnarray}
	\frac{{\rm d}N}{{\rm d}\phi} = \frac{1}{2\pi}\Bigg[1+\sum_{n=1}^{\infty}2v_n \cos{\left(n(\phi-\psi_n)\right)}\Bigg].
	\label{vn}
\end{eqnarray}
The $n^{\rm th}$ order of harmonic flow coefficient $v_n = \langle \cos{[n(\phi-\psi_n)]}\rangle$ characterizes the magnitude of azimuthal anisotropies of the particle spectrum in the transverse directions~\cite{Voloshin:1994mz}, while the phase $\psi_n$ is the $n^{\rm th}$ harmonic event plane angle. The second ($v_2$) and third ($v_3$) Fourier coefficients represent the amplitudes of elliptic and triangular flow, respectively. The linearized hydrodynamic response shows that anisotropy flow $v_n$ is likely correlated to the geometry asymmetry of energy density profile in spatial space of the initial state, namely the initial eccentricity $\varepsilon_n$,
\begin{equation}
	\varepsilon_n=\frac{\sqrt{\langle r^n\cos(n\phi_r)\rangle^2+\langle r^n \sin(n\phi_r)\rangle^2}}{\langle r^n \rangle},
	\label{epsilon_n}
\end{equation}
where $r$ and $\phi_r$ are the polar coordinates of participating nucleons~\cite{Alver:2010gr}. In hydrodynamics, harmonic flows are responding to eccentricities,
\begin{equation}
	v_n=v_n(\varepsilon_n, k),
	\label{relatation}
\end{equation}
where the constant $k$ is sensitive to the properties of the QGP, such as the transport coefficient $\eta/s$~\cite{Yan:2017ivm}.

Note that the above definition of $v_n$ according to $\psi_n$ is based on the assumptions that the QGP is created and governed by hydrodynamics in relativistic heavy-ion collisions. This has been well demonstrated in large colliding systems, but needs to be clarified in small systems since whether small colliding systems can create the QGP is still debatable. To find differences in the sources and characteristics of collective flow between large and small colliding systems, we will consider the PCN as a new approach to study different EbyE features of collective behavior between small and large colliding systems.

\section{Training the PCN for classifying two systems} 
 \label{PCN}
 In this section, we introduce the detailed analysis procedures, including the PCN architecture, input, output, training, and evaluation of the PCN.  

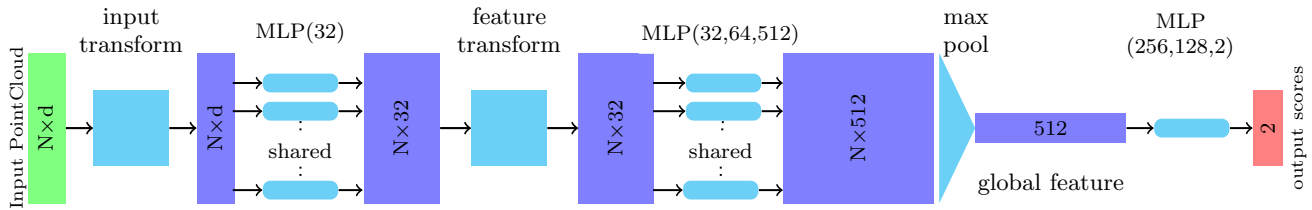
\begin{figure*}[hbtp]
	\begin{center}
		\begin{tikzpicture}[scale=0.75]
			\node [rectangle,fill=red!0,rotate=90] (in) at (0.0,3) {\scriptsize{Input PointCloud}};			
			\node [rectangle,fill=green!50,rotate=90,minimum width=2cm, minimum height=0.5cm] (input) at (0.5,3) {\footnotesize{N$\times$d}};
			
			\node (text) at (2,4.7) {$\begin{matrix} \text{input}\\\text{transform} \end{matrix}$};
			\node [rectangle,fill=cyan!50,rotate=90,minimum width=1cm, minimum height=1cm] (transform1) at (2,3) {};
			\draw [->,color=black,line width=0.8pt] (input)-- (transform1);
			
			\node [rectangle,fill=blue!50,rotate=90,minimum width=2cm, minimum height=0.5cm] (input1) at (3.5,3) {\footnotesize{N$\times$d}};
			\draw [->,color=black,line width=0.8pt] (transform1)-- (input1);

			\coordinate (po1-1) at (3.8,4.5);
			\coordinate (po1-2) at (3.8,3.8);
			\coordinate (po1-3) at (3.8,3.3);
			\coordinate (po1-4) at (3.8,2.6);
			\coordinate (po1-5) at (3.8,1.9);
			
			\node [rectangle,rounded corners=1mm,fill=cyan!0,minimum width=1cm, minimum height=0.25cm] (mlp1) at (5,4.7) {\footnotesize{MLP(32)}};
			\node [rectangle,rounded corners=1mm, fill=cyan!50,minimum width=1cm, minimum height=0.25cm] (mlp2) at (5,3.8) {};
			\node [rectangle,rounded corners=1mm,fill=cyan!50,minimum width=1cm, minimum height=0.25cm] (mlp3) at (5,3.3) {};
			\node [rectangle,rounded corners=1mm,fill=cyan!0,minimum width=1cm, minimum height=0.25cm] (mlp4) at (5,2.6) {\footnotesize{shared}};
			\node [rectangle,rounded corners=1mm,fill=cyan!50,minimum width=1cm, minimum height=0.25cm] (mlp5) at (5,1.9) {};
			
			\draw [->,color=black,line width=0.75pt] (po1-2)-- (mlp2);
			\draw [->,color=black,line width=0.75pt] (po1-3)-- (mlp3);
			\draw [->,color=black,line width=0.75pt] (po1-5)-- (mlp5);
			\draw [dotted,color=black,line width=0.75pt] (mlp3)-- (mlp4);
			\draw [dotted,color=black,line width=0.75pt] (mlp4)-- (mlp5);
			
			\coordinate (po2-1) at (6.1,4.5);
			\coordinate (po2-2) at (6.1,3.8);
			\coordinate (po2-3) at (6.1,3.3);
			\coordinate (po2-4) at (6.1,2.6);
			\coordinate (po2-5) at (6.1,1.9);
			
			\draw [<-,color=black,line width=0.75pt] (po2-2)-- (mlp2);
			\draw [<-,color=black,line width=0.75pt] (po2-3)-- (mlp3);
			\draw [<-,color=black,line width=0.75pt] (po2-5)-- (mlp5);
			\node [rectangle,fill=blue!50,rotate=90,minimum width=2cm, minimum height=1cm] (feature1) at (6.8,3) {\footnotesize{N$\times$32}};

			\node (text1) at (8.7,4.7) {$\begin{matrix} 		\text{feature}\\\text{transform} \end{matrix}$};
			\node [rectangle,fill=cyan!50,rotate=90,minimum width=1cm, minimum height=1cm] (transform2) at (8.7,3) {};
			\draw [->,color=black,line width=0.8pt] (feature1)-- (transform2);
			
			\node [rectangle,fill=blue!50,rotate=90,minimum width=2cm, minimum height=1cm] (feature2) at (10.6,3) {\footnotesize{N$\times$32}};
			\draw [->,color=black,line width=0.8pt] (transform2)-- (feature2);
			
			\coordinate (po3-1) at (11.25,4.5);
			\coordinate (po3-2) at (11.25,3.8);
			\coordinate (po3-3) at (11.25,3.3);
			\coordinate (po3-4) at (11.25,2.6);
			\coordinate (po3-5) at (11.25,1.9);
			
			\node [rectangle,rounded corners=1mm,fill=cyan!0,minimum width=1cm, minimum height=0.25cm] (mlp21) at (12.45,4.65) {\footnotesize{MLP(32,64,512)}};
			\node [rectangle,rounded corners=1mm, fill=cyan!50,minimum width=1cm, minimum height=0.25cm] (mlp22) at (12.5,3.8) {};
			\node [rectangle,rounded corners=1mm,fill=cyan!50,minimum width=1cm, minimum height=0.25cm] (mlp23) at (12.5,3.3) {};
			\node [rectangle,rounded corners=1mm,fill=cyan!0,minimum width=1cm, minimum height=0.25cm] (mlp24) at (12.45,2.6) {\footnotesize{shared}};
			\node [rectangle,rounded corners=1mm,fill=cyan!50,minimum width=1cm, minimum height=0.25cm] (mlp25) at (12.45,1.9) {};
			
			\draw [->,color=black,line width=0.75pt] (po3-2)-- (mlp22);
			\draw [->,color=black,line width=0.75pt] (po3-3)-- (mlp23);
			\draw [->,color=black,line width=0.75pt] (po3-5)-- (mlp25);
			\draw [dotted,color=black,line width=0.75pt] (mlp23)-- (mlp24);
			\draw [dotted,color=black,line width=0.75pt] (mlp24)-- (mlp25);
			
			\coordinate (po4-1) at (13.55,4.5);
			\coordinate (po4-2) at (13.55,3.8);
			\coordinate (po4-3) at (13.55,3.3);
			\coordinate (po4-4) at (13.55,2.6);
			\coordinate (po4-5) at (13.55,1.9);
			
			\draw [<-,color=black,line width=0.75pt] (po4-2)-- (mlp22);
			\draw [<-,color=black,line width=0.75pt] (po4-3)-- (mlp23);
			\draw [<-,color=black,line width=0.75pt] (po4-5)-- (mlp25);
			\node [rectangle,fill=blue!50,rotate=90,minimum width=2cm, minimum height=2.cm] (feature3) at (14.9,3) {\footnotesize{N$\times$512}};
			
			\node [isosceles triangle,fill=cyan!50, isosceles triangle stretches, minimum height=0.05cm, minimum width=2.cm] (feature4) at (16.6,3) {};

			\node (text3) at (16.8,4.7) {$\begin{matrix} 		\text{max}\\\text{pool} \end{matrix}$};
			\node [rectangle,fill=blue!50,minimum width=2cm, minimum height=0.1cm] (feature4) at (18.3,3) {\footnotesize{512}};
			\node [rectangle,rounded corners=1mm,fill=cyan!0,minimum width=1cm, minimum height=0.25cm] (mlp31) at (20.6,4.65) {\footnotesize{$\begin{matrix}\text{MLP} \\ \text{(256,128,2)}\end{matrix}$}};
			\node [rectangle,rounded corners=1mm,fill=cyan!50,minimum width=1cm, minimum height=0.25cm] (mlp32) at (20.8,3) {};
			
			\node [rectangle,rounded corners=1mm,fill=cyan!0,minimum width=1cm, minimum height=0.25cm] (text4) at (18.3,2) {global feature};
			
			\draw [->,color=black,line width=0.75pt] (feature4)-- (mlp32);
			
			\node [rectangle,fill=red!50,minimum width=1cm, minimum height=0.2cm,rotate=90] (o3) at (22.15,3) {\footnotesize{2}};
			
			\draw [->,color=black,line width=0.75pt] (mlp32)-- (o3);
			\node [rectangle,fill=red!0,rotate=90] (out) at (22.7,3) {\scriptsize{output scores}};
			
		\end{tikzpicture}
	\end{center}
	\caption{The architecture of the devised point cloud network (PCN). In this network, it takes final state $N$ points as input, applies input and feature transformations, and then aggregates point features by max-pooling. The output is classification scores for two classes, i.e., Pb $+$ Pb or $p$ $+$ Pb collision.}
	\label{myNet}
\end{figure*}
\subsection{Network architecture}
\label{Architecture}

The architecture of the PCN is shown in Fig.\ref{myNet}. It begins with an input alignment network , which initiates the process of aligning the particle clouds in input space, and also enables the model to capture correlations irrespective of orientation in input space. The following is a shared pointwise multilayer perceptron (MLP)\footnote{An MLP is a type of feedforward neural network composed of multiple layers of nodes in a directed graph. Here in our model the pointwise MLP refers to an MLP that processes each point from the input independently.} implemented by a 1D-convolution neural network (CNN)\footnote{1D CNNs are a type of neural network that uses 1D convolution operations to process and transform their input data.} to extract 32 feature maps\footnote{A feature map, in the context of convolutional neural network in particular, is the output of one convolutional kernel (i.e., filter) applied to the previous layer, which can capture specific patterns or features from the input data, highlighting areas that match the convolutional kernel's pattern.}, a feature alignment network, a shared MLP to extract 32, 64 and 512 feature maps, respectively. A global max pooling\footnote{A global max pooling operation is retrieving the maximum values of each feature across all particles, summarizing them as a singular global feature representative of the entire particle cloud.} then gets the maximum values of each feature among all particles as one global feature of the particle cloud. Finally, a shared MLP implemented by three layers fully connected deep neural network (DNN)\footnote{A fully connected DNN is a neural network architecture where each node(i.e.,neuron) in a layer is connected to every node in the subsequent layer, facilitating complex pattern recognition through multiple layers of computation.} with 256, 128, and 2 neurons tags each event as $p$ $+$ Pb or Pb $+$ Pb collision. Batch normalization\footnote{Batch normalization is a technique in deep learning that normalizes the inputs of each layer, to improve training stability and performance by reducing internal covariate shift, for more detail see Ref.\cite{2015arXiv150203167I}.} layers are present between every convolution layer. The LeakyReLU($\alpha$=0.01) activation function\footnote{The LeakyReLU activation function is defined as $f(x)=x$ if $x>0$ and $f(x)=\alpha x$ otherwise, where $\alpha$ is a small, positive parameter.} is used for all layers except the final layer. The sigmoid activation [$\sigma(x)=1/(1+e^{-x})$] is used on the final layer for binary classification. The models use the Adam optimizer\cite{2014arXiv1412.6980K} with a learning rate of $10^{-4}$ with total decay  $10^{-4}$ and categorical cross entropy as the loss function. In addition, a dropout\cite{JMLR:v15:srivastava14a} layers (with drop out probability 0.3) and L2 regularization\cite{10.1145/1015330.1015435} are present to tackle the overfitting issue. We use a maximum of 50 epochs with 32 batch size to train the data set. The architecture of the PCN we use is similar to the original architecture as described in Ref.\cite{Qi:2017pn} but less complex. The choice of specific hyperparameters, including the number of nodes and layers, was driven by a combination of empirical experimentation and previous successful implementations in analyzing heavy-ion collision data. Our primary goal is to investigate the identifiability with machine learning perspective of the collective flow characteristics in $p$ $+$ Pb or Pb $+$ Pb collisions. Point clouds of particles are processed with transformations order invariant operations to extract global features. While a fully connected deep neural network (DNN) tags each event as $p$ $+$ Pb or Pb $+$ Pb collision.

 \subsection{Input and output of the machine learning}
Following the CMS method as described in Ref.\cite{CMS:2013jlh}{}, the AMPT events for both $p$ $+$ Pb and Pb $+$ Pb collisions are grouped into the data sets according to the number of final charged particles $N_{ch}$ measured in the kinetic window of $|\eta| <$ 2.4 and $p_{\rm T} >$ 0.4 GeV/\textit{c}. These data are divided into various centrality classes for the colliding systems, which are the $N_{ch}$ bins of 90-120, 120-150, 150-185, 185-230, to compare different systems at similar volumes. Each data set has about 0.3 million events, which are divided into training events, validation events, and test events by the ratio of 5:2:3. The value and error bar of test accuracy are the mean value and the standard deviation of the test data set divided into 100 pieces, respectively. All the information about the final state particles within $|\eta| <$ 2.4 in each event is the input to the PCN as a sample, which consists of a list of particles with their information on ($p_{\rm x}$, $p_{\rm y}$, $p_z$, $E$) or ($p_{\rm x}$, $p_{\rm y}$).  Simultaneously, the two true labels, $p$ $+$ Pb and Pb$+$Pb, are marked on each event to perform the supervised training. 

\section{Training results and discussion}
\label{results}
In our study, two cases of input data have been investigated. In the case 1, the input for training is an EbyE list of four-momentum ($p_{\rm x}$, $p_{\rm y}$, $p_z$, $E$) of the selected final state hadrons from the AMPT model. The input data have a dimension of $N\times4$, where $N$ is the maximum number of selected particles in an event. Events with fewer particles are filled with zeros to maintain the same input dimension. The case 2 is the same as the case 1, but the input is a list of two-momentum ($p_{\rm x}$, $p_{\rm y}$) of selected final state hadrons. 

 \subsection{Case 1: training with four-momentum of final hadrons}
\label{four}
\begin{figure}[hbtp]
	\includegraphics[scale=0.40]{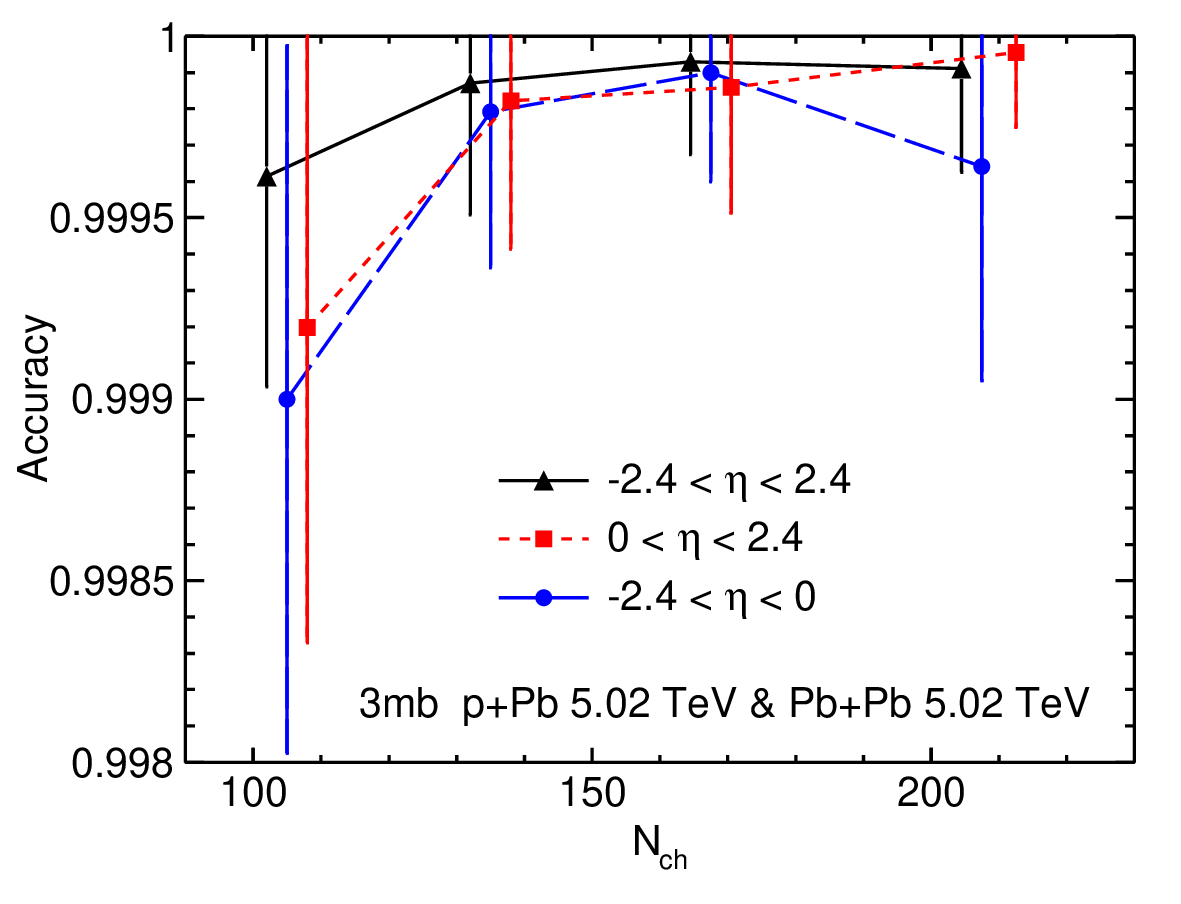}
	\caption{(Color online) Test accuracy as a function of the number of charged particles $N_{ch}$ by training with four-momentum ($p_{\rm x}$, $p_{\rm y}$, $p_z$, $E$) of final particles in pseudorapidity ranges -2.4$<\eta<$2.4 (solid curve), 0$<\eta<$2.4 (dashed curve), and -2.4$<\eta<$0 (long-dashed curve). }
	\label{train4p}
\end{figure}

Figure~\ref{train4p} shows test accuracy as a function of the number of charged particles $N_{ch}$ by learning four-momentum ($p_{\rm x}$, $p_{\rm y}$, $p_z$, $E$) of final particles in three different pseudorapidity ranges. All the test accuracies are higher than 99\%, which indicates that the two colliding systems are sufficiently distinguished by training the input of four-momentum ($p_{\rm x}$, $p_{\rm y}$, $p_z$, $E$) of final particles.

\begin{figure}[hbtp]
	\includegraphics[scale=0.40]{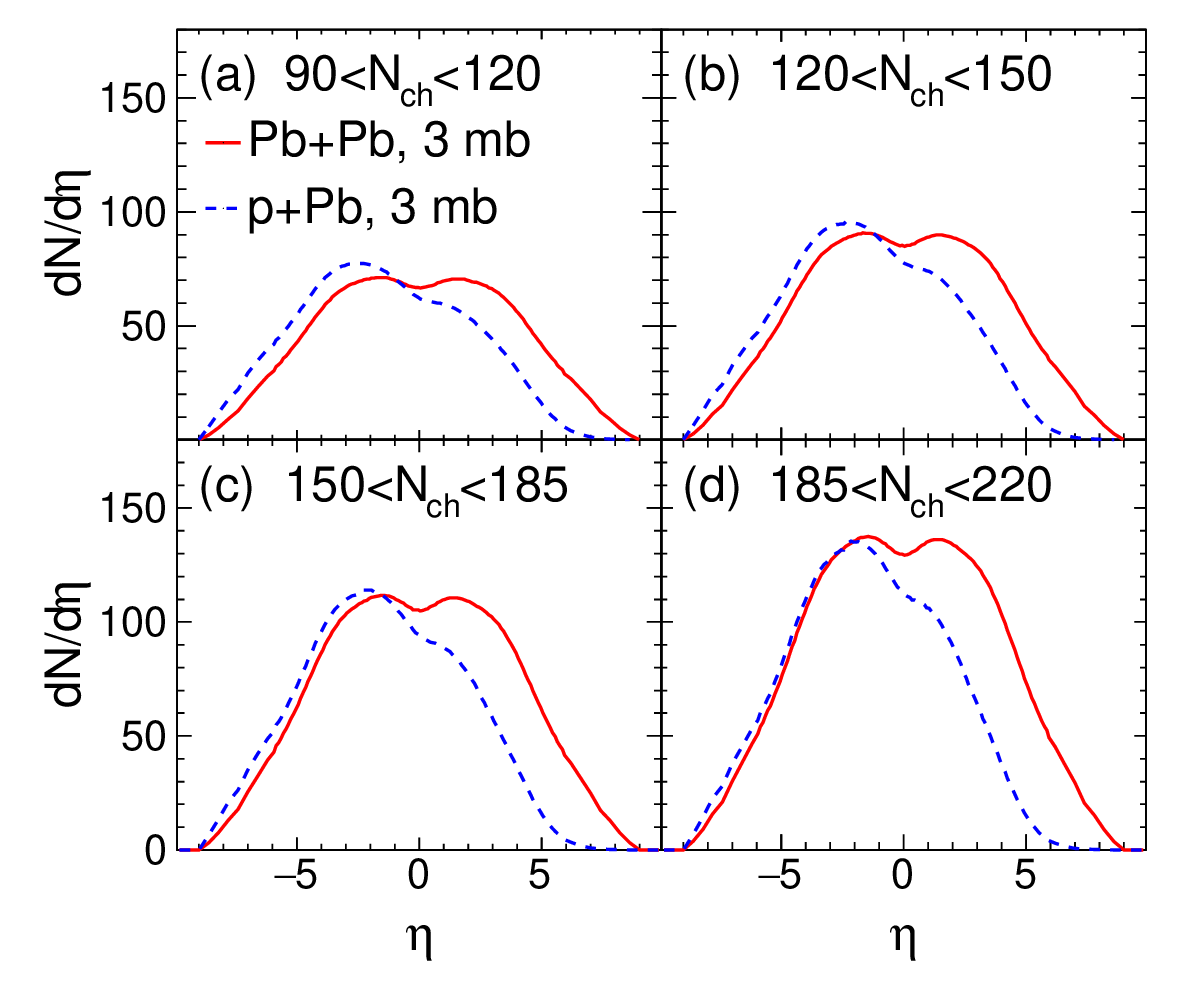}
	\caption{(Color online) The pseudorapidity distributions of final hadrons for different $N_{ch}$ classes in Pb$ + $Pb collisions (solid curve) and $p$ $+$ Pb collisions (dashed curve).}
	\label{dndy}
\end{figure}

Final particles in three pseudorapidity ranges (-2.4$<\eta<$2.4, -2.4$<\eta<$0 and 0$<\eta<$2.4) were used to train, due to the different pseudorapidity distributions between Pb $+$ Pb collisions and $p$ $+$ Pb collisions, which is shown in Fig.~\ref{dndy}. The lower hadron yield in the $p$ -going (forward) direction for $p$ $+$ Pb collisions was expected to induce the sensitivity to the test accuracy of the pseudorapidity range. However, in any given circumstance, the accuracy is high enough with the margin of error, which demonstrates the PCN is capable of identifying two colliding systems on an event-by-event basis, despite almost the same distribution between the two colliding systems in backward pseudorapidity range.

 \subsection{Case 2: training with two-momentum of final hadrons}
\label{two}

To investigate whether the PCN can learn the difference of collective flow between $p+$Pb collisions and peripheral Pb $+$ Pb collisions, three scenarios of lists are used as training inputs, i.e., two-momentum ($p_{\rm x}$, $p_{\rm y}$), normalized two-momentum ($p_{\rm x}^{\rm norm}$, $p_{\rm y}^{\rm norm}$), and normalized meanwhile randomly rotated two-momentum ($p_{\rm x}^{\rm norm, rand}$, $p_{\rm y}^{\rm norm, rand}$), respectively. The normalized two-momentum $p_{\rm x}^{\rm norm}$ and $p_{\rm y}^{\rm norm}$ are defined as,
\begin{eqnarray}\label{pt_norm}
p_{\rm x}^{\rm norm} & = & \frac{p_{\rm x}}{p_{\rm T}},  \notag \\
p_{\rm y}^{\rm norm} & = & \frac{p_{\rm y}}{p_{\rm T}},
\end{eqnarray}
where $p_{\rm T}$ is the transverse momentum of each particle. The randomly rotated two-momentum $p_{\rm x}^{\rm rand}$ and $p_{\rm y}^{\rm rand}$ are,
\begin{eqnarray}\label{pt_rand}
	p_{\rm x}^{\rm rand} & = & p_{\rm x}\times\cos{\phi_{\rm rand}}-p_{\rm y}\times\sin{\phi_{\rm rand}},  \notag \\
	p_{\rm y}^{\rm rand} & = & p_{\rm x}\times\sin{\phi_{\rm rand}}-p_{\rm y}\times\cos{\phi_{\rm rand}},
\end{eqnarray}
where $\phi_{\rm rand}$ is a random angle between 0 and 2$\pi$. While, the normalized and randomly rotated two-momentum $p_{\rm x}^{\rm norm, rand}$ and $p_{\rm y}^{\rm norm, rand}$ are,
\begin{eqnarray}\label{pt_norm_rand}
	p_{\rm x}^{\rm norm, rand} & = & \frac{p_{\rm x}\times\cos{\phi_{\rm rand}}-p_{\rm y}\times\sin{\phi_{\rm rand}}}{p_{\rm T}},  \notag \\
	p_{\rm y}^{\rm norm, rand} & = & \frac{p_{\rm x}\times\sin{\phi_{\rm rand}}-p_{\rm y}\times\cos{\phi_{\rm rand}}}{p_{\rm T}}.
\end{eqnarray}

\begin{figure}[hbtp]
	\includegraphics[scale=0.45]{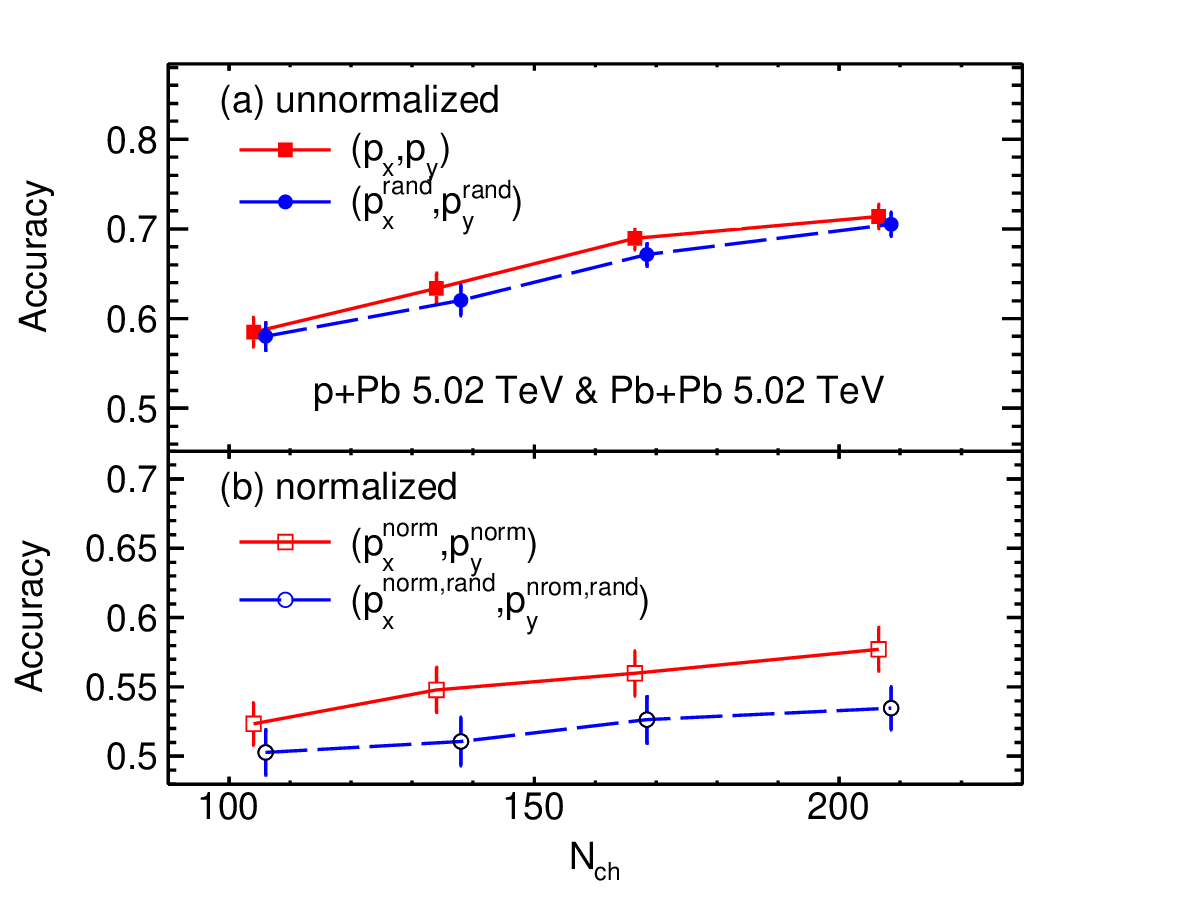}
	\caption{(Color online) Test accuracy as a function of the number of charged particles $N_{ch}$ by learning with input to be two-momentum ($p_{\rm x}$, $p_{\rm y}$) (a) and the normalized two-momentum ($p_{\rm x}^{\rm norm}$, $p_{\rm y}^{\rm norm}$) (b), without (solid curve) or with (dashed curve) random rotations.}
	\label{norm_rand}
\end{figure}

Figure \ref{norm_rand}(a) shows test accuracy as a function of the number of final charged particles $N_{ch}$ with input to be two-momentum ($p_{\rm x}$, $p_{\rm y}$) and its random rotation ($p_{\rm x}^{\rm rand}$ and $p_{\rm y}^{\rm rand}$) of final particles. Test accuracy of two-momentum input is over $60\%$, and it will drop less than $2\%$, if two-momentum are randomly rotated. It indicates that besides anisotropic flow there is other discrepancy between the two colliding systems, like in $p_{\rm T}$ distributions. To eliminate the effect from different $p_{\rm T}$ distribution of hadron, we train the PCN with normalized two-momentum ($p_{\rm x}^{\rm norm}$, $p_{\rm y}^{\rm norm}$) and its random rotation ($p_{\rm x}^{\rm norm, rand}$ and $p_{\rm y}^{\rm norm, rand}$) of final particles. The test accuracy as a function of the number of final charged particles is shown in Fig.\ref{norm_rand}(b). Compared to the normal two-momentum training, there is a more than 10\% decrease in test accuracy after normalization. This indicates that $p_{\rm T}$ distribution of hadrons is the main feature that distinguishes the two colliding systems, if we train the PCN by two-momentum. Furthermore, the PCN can not distinguish the two colliding systems by learning normalized two-momentum with random rotations of final hadrons, because these operations eliminate the information about the magnitude of $p_{\rm T}$ and anisotropic flow. Compared to the training by the normalized two-momentum with random rotations, the test accuracy is improved by a few percentages by learning two-momentum with the normalization only. This discrepancy comes from the difference in anisotropic flow between two colliding systems. However, because the anisotropic flows between two colliding systems are very similar, the PCN can not distinguish the two different colliding systems very well, even though it is better than the case with random rotations.

\begin{figure}[hbtp]
	\includegraphics[scale=0.40]{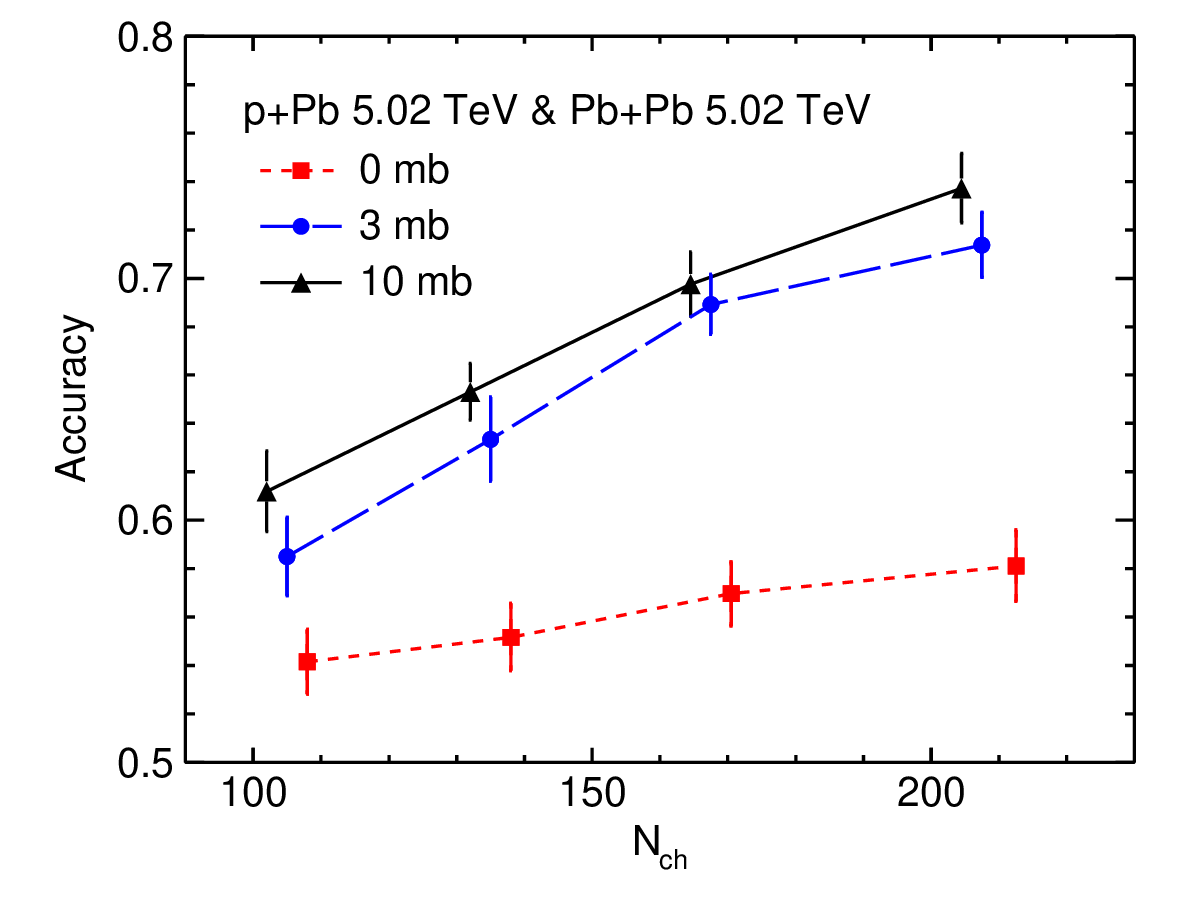}
	\caption{(Color online) Test accuracy as a function of the number of charged particles $N_{ch}$ by learning two-momentum ($p_{\rm x}$, $p_{\rm y}$) of final hadrons in the AMPT model with different parton cross sections.}
	\label{ACS_pt}
\end{figure}

\begin{figure}[hbtp]
	\includegraphics[scale=0.4]{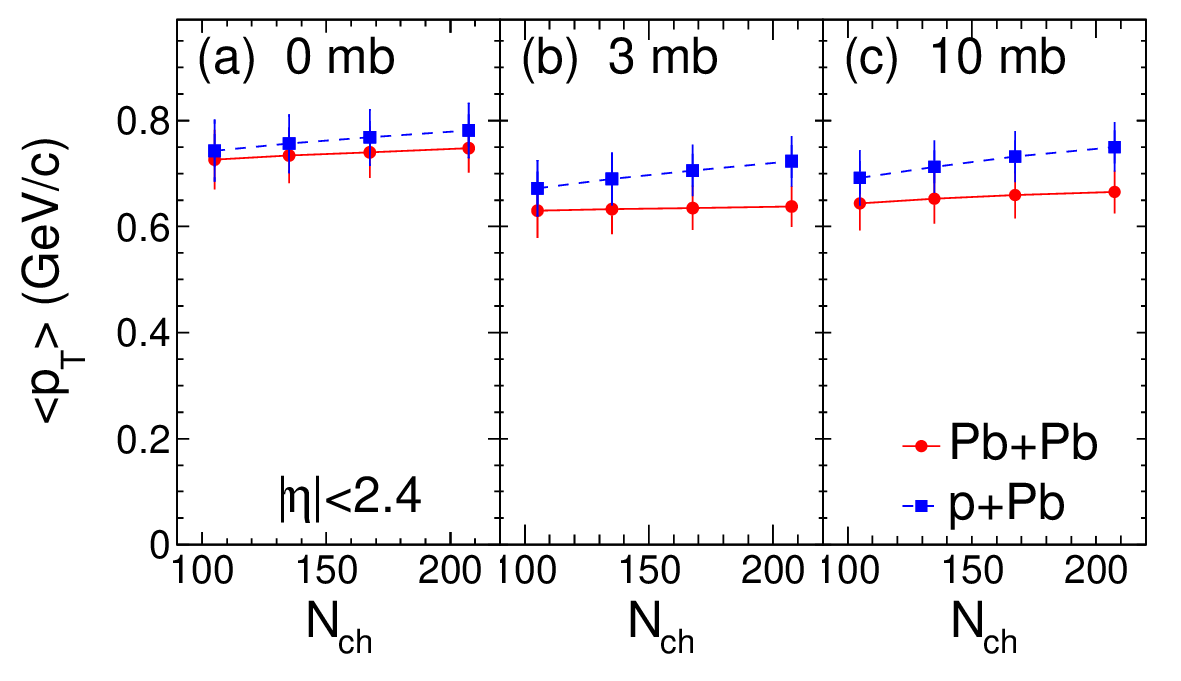}
	\caption{(Color online) Average transverse momenta $\left\langle p_{\rm T} \right\rangle $ in Pb $+$ Pb collisions (solid curve) and p $+$ Pb collisions (dashed curve) as a function of the number of charged particles $N_{ch}$ with parton cross section of 0 mb (a), 3 mb (b), and 10 mb (c).}
	\label{meanpt}
\end{figure}

Figure~\ref{ACS_pt} shows the test accuracy as a function of the number of final charged particles $N_{ch}$ by learning two-momentum ($p_{\rm x}$, $p_{\rm y}$) of final particles with different parton cross sections. It can be seen that all test accuracies increase with the increase of  $N_{ch}$ and parton cross section. In addition, the PCN can not distinguish the two colliding systems with 0 mb parton cross section at low $N_{ch}$. These results can basically be explained by Fig.~\ref{meanpt}, since the average transverse momentum $\left\langle p_{\rm T} \right\rangle $ can quantify the feature of $p_{\rm T}$ spectra shape.

Figure \ref{meanpt} shows average transverse momenta $\left\langle p_{\rm T} \right\rangle $ as a function of the number of final charged particles $N_{ch}$ with different parton cross sections. We can see that the $\left\langle p_{\rm T} \right\rangle $ in $p$ $+$ Pb collisions is larger than that in Pb $+$ Pb collisions, and the difference is more significant with a larger parton cross section and higher multiplicity. 
Although we observe that the difference of $\left\langle p_{\rm T} \right\rangle $ is almost zero for 0 mb parton cross section, some test accuracies are larger than 0.5 for 0 mb in Fig.~\ref{ACS_pt}. It indicates that there is a difference in non-flow effect between small and large colliding systems. It could be attributed to jets, because the impact of the jet transverse momentum broadening and multiple scatterings~\cite{Leonidov:1997iz} has been found to be stronger in $p$ $+$ Pb collisions than in peripheral Pb $+$ Pb collisions~\cite{STAR:2008med}.

\begin{figure}[hbtp]
	\includegraphics[scale=0.40]{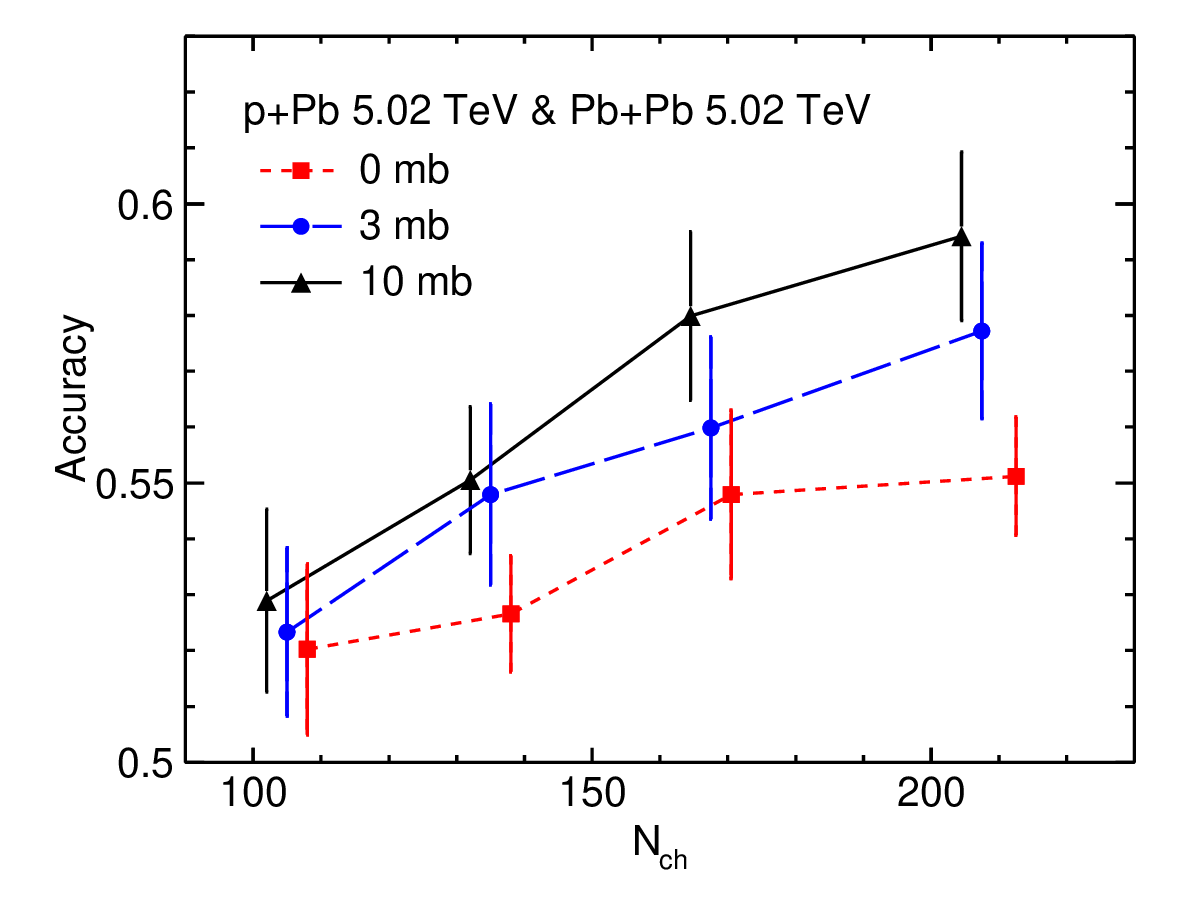}
	\caption{(Color online) Test accuracy as a function of the number of charged particles $N_{ch}$ by learning normalized two-momentum ($p_{\rm x}^{\rm norm}$, $p_{\rm y}^{\rm norm}$) of final hadrons in the AMPT model with different parton cross sections.}
	\label{ACS}
\end{figure}

Figure~\ref{ACS} shows test accuracy as a function of the number of final charged particles $N_{ch}$ by learning normalized two-momentum ($p_{\rm x}^{\rm norm}$, $p_{\rm y}^{\rm norm}$) of final particles with different parton cross sections. 
It can be seen that the test accuracy with 0 mb is close to the corresponding result in Fig.~\ref{ACS_pt}, and both of them increase with $N_{ch}$, which indicates that the discrepancy of non-flow effect between the two systems is larger for higher multiplicity. However, the discrepancy between 0 mb and nonzero parton cross sections must come from anisotropic flow. The discrepancy increases with parton cross section, indicating a more significant discrepancy of anisotropic flow between two colliding systems due to more parton scatterings.

\begin{figure}[hbtp]
	\includegraphics[scale=0.4]{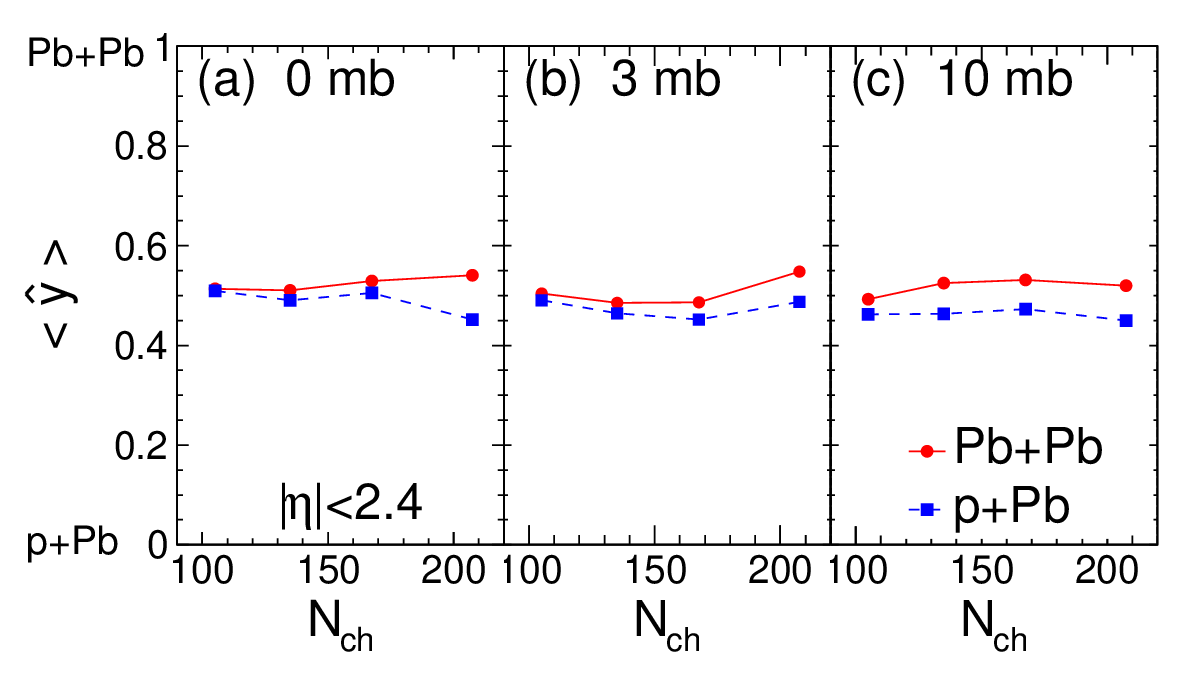}
	\caption{(Color online) Averaged output $\left\langle \widehat{y} \right\rangle$ of the ensembles of two systems identified by the model trained by normalized two-momentum ($p_{\rm x}^{\rm norm}$, $p_{\rm y}^{\rm norm}$) of final hadrons in the AMPT model with different parton cross sections.}
	\label{ensemble}
\end{figure}
If an event is identified as $p$ $+$ Pb or Pb $+$ Pb, it is marked by the output of $\widehat{y}$=0 or 1, respectively, in our analysis. Thus, the averaged output $\left\langle \widehat{y} \right\rangle$ of our model can represent the probability of Pb $+$ Pb of an event. In other words, The closer the $\left\langle \widehat{y} \right\rangle$ is to 0, the more likely the events are $p$ $+$ Pb collisions, and vice versa for Pb $+$ Pb collisions. Figure~\ref{ensemble} shows the averaged output $\left\langle \widehat{y} \right\rangle$ of the ensembles of each system identified by the model trained by normalized two-momentum ($p_{\rm x}^{\rm norm}$, $p_{\rm y}^{\rm norm}$) of final hadrons in the AMPT model with different parton cross sections. It can be seen that $\left\langle \widehat{y} \right\rangle$ of the ensembles of Pb $+$ Pb events is larger than that for $p$ $+$ Pb events in every situation, which indicates that the PCN is able to distinguish two different ensembles of systems, according to the difference of averaged anisotropic flow between two systems. 

To investigate the relationship between anisotropic flow and parton cross section, we further calculate the harmonic flow coefficients, e.g., elliptic flow $v_2$, by fitting the long-range part of the two-particle azimuthal correlation function $C(\Delta\phi)$, which defined as
\begin{equation}
	C(\Delta\phi)=\frac{Y_{same}(\Delta\phi)}{Y_{mixed}(\Delta\phi)} \times \frac{\int Y_{mixed}(\Delta\phi)d\Delta\phi}{\int Y_{same}(\Delta\phi)d\Delta\phi},
	\label{correlation}
\end{equation}
where $Y_{same}(\Delta\phi=\phi_2-\phi_1)$ and $Y_{mixed}(\Delta\phi)$ are the number of particle pairs 
at a given $\Delta\phi$ and within a given $p_{\rm T}$ range for the same and mixed events. This definition of  $C(\Delta\phi)$ removes a trivial dependence on the number of produced particles
\cite{PHENIX:2005zfm,PHENIX:2014fnc,Bzdak:2014dia}.

\begin{figure}[hbtp]
	\includegraphics[scale=0.40]{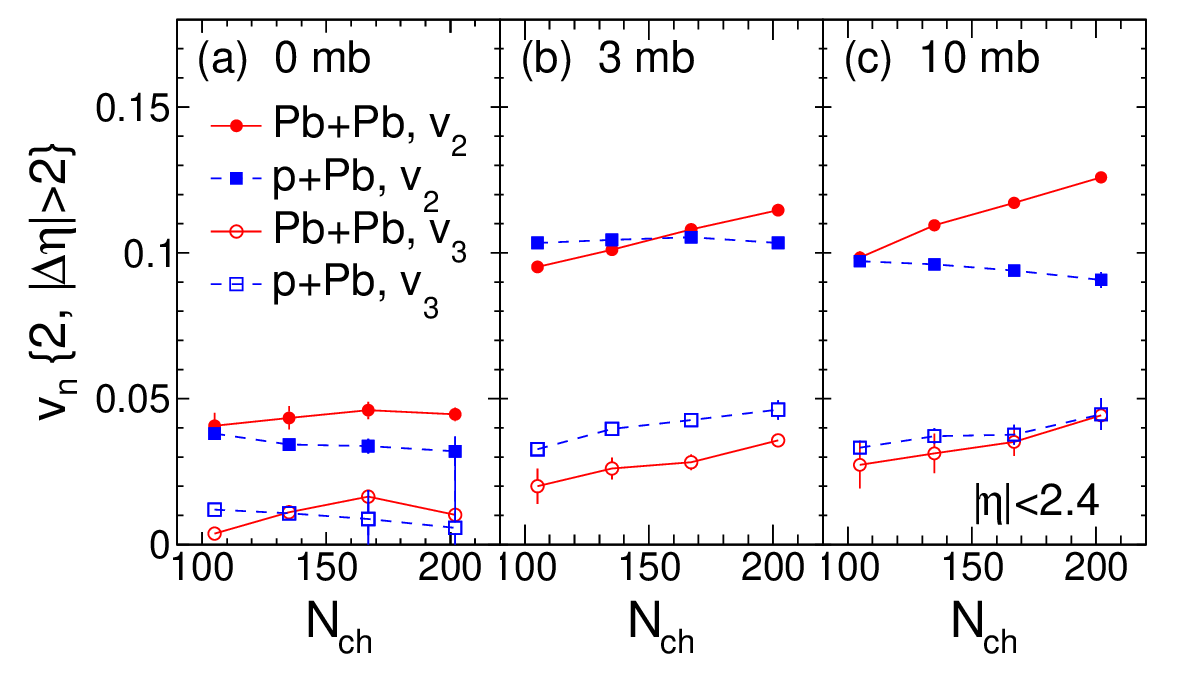}
	\caption{(Color online) Integrated elliptic  $v_2$ and triangular $v_3$ flow coefficient in Pb $+$ Pb collisions (solid curve) and p $+$ Pb collisions (dashed curve) as a function of the number of charged particles $N_{ch}$ with parton cross section of 0 mb (a), 3 mb (b), and 10 mb (c).}
	\label{vn}
\end{figure}

Figure \ref{vn} shows $v_2$ and  $v_3$ as a function of the number of final charged particles $N_{ch}$ 
with different parton cross sections. It can be observed that $v_2$ and $v_3$ increases with parton cross section for Pb $+$ Pb collisions. But the dependence on parton cross section is non-monotonous for $p$ $+$ Pb collisions, which has already been found in Ref.~\cite{Zhao:2021bef}. The most significant difference of $v_2$ between the two systems appears when parton cross section is taken as 10 mb for high-multiplicity events, because there are the largest $v_2$ in Pb $+$ Pb collisions and relatively small $v_2$ in $p$ $+$ Pb collisions. On the other hand, $v_3$ of the two systems are similar for the two parton cross sections of 0 mb and 10 mb. However, there is a more obvious difference of $v_3$ between the two systems with parton cross section of 3 mb, because there are the largest $v_3$ in $p$ $+$ Pb collisions and relatively small $v_3$ in Pb $+$ Pb collisions. Surprisingly, the discrepancy of $v_2$ and $v_3$ between two ensembles of each system can be captured by the PCN, even the similar $N_{ch}$ and parton cross section dependences are obtained in Figs.~\ref{ACS} and ~\ref{ensemble}, although the test accuracies are not high enough for the PCN to distinguish two systems due to similar EbyE flow distributions which will be shown next. 

\begin{figure*}[hbtp]
	\includegraphics[scale=0.70]{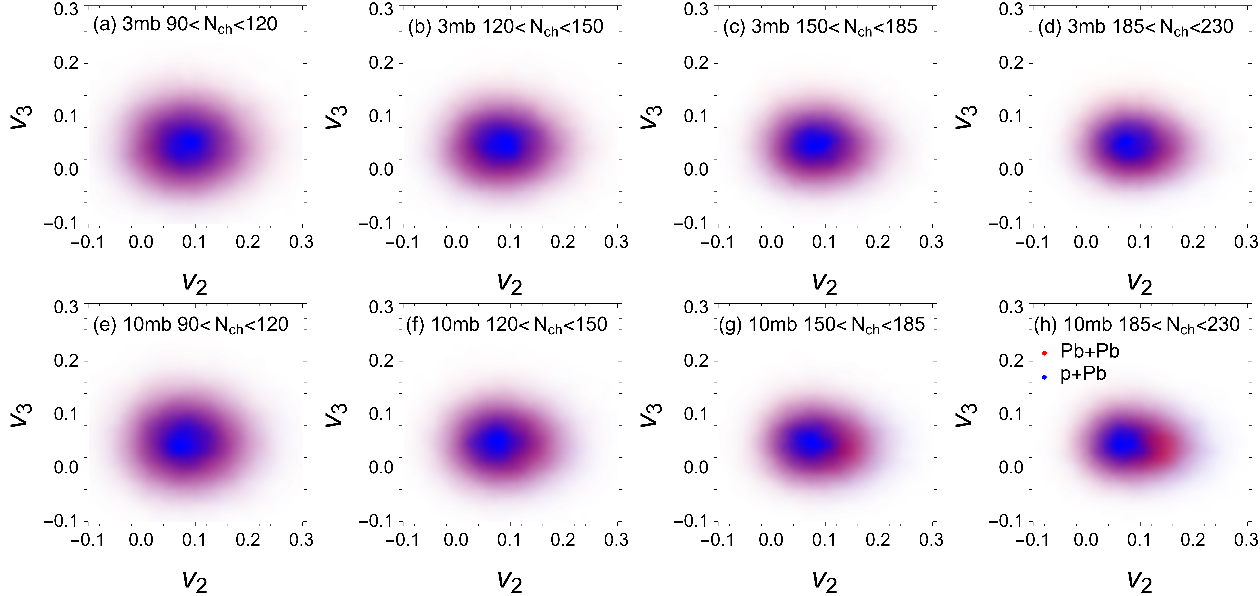}
	\caption{(Color online) The EbyE 2-D distributions of $v_2$ vs $v_3$ [P($v_2$, $v_3$)] of Pb $+$ Pb (red) and $p$ $+$ Pb (blue) collisions in the AMPT model with the parton cross section of 3 mb (first row) and 10 mb (second row) for the $N_{ch}$ bin of 90$<N_{ch}<$120 (first column), 120$<N_{ch}<$150 (second column), 150$<N_{ch}<$185 (third column), and 185$<N_{ch}<$230 (fourth column).}
	\label{distr}
\end{figure*}

\begin{figure}[hbtp]
	\includegraphics[scale=0.4]{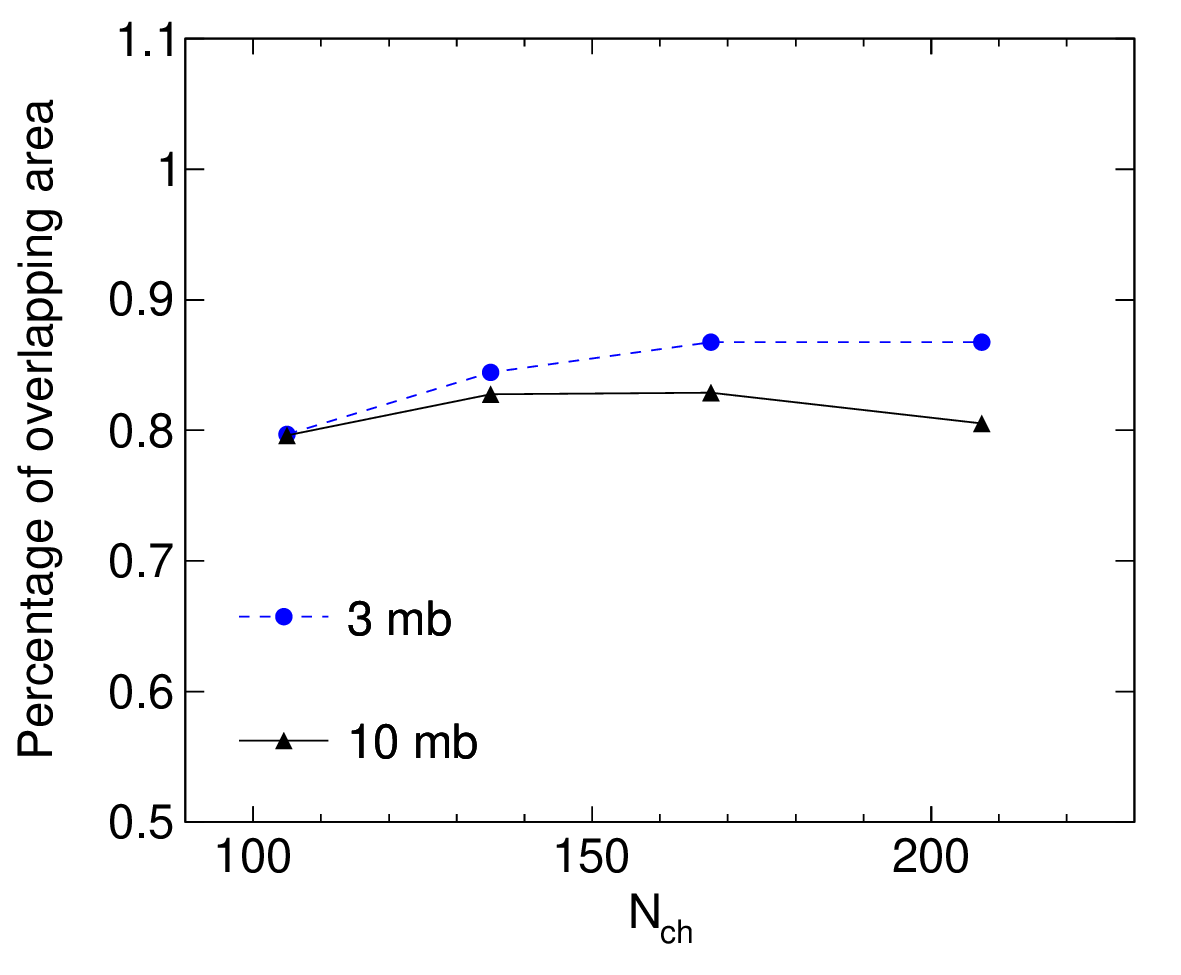}
	\caption{(Color online) The percentage of the 2-D overlapping volume between the distribution P($v_2$,$v_3$) for Pb $+$ Pb collisions and that for $p$ $+$ Pb collisions as a function of the number of final charged particles $N_{ch}$ in the AMPT model with the parton cross section of 0 mb (solid curve), 3 mb (dashed curve), and 10 mb (dotted curve).}
	\label{overlap}	
\end{figure}

Figure \ref{distr} shows the EbyE 2-D distributions of $v_2$ vs $v_3$ [P($v_2$, $v_3$)] in Pb $+$ Pb collisions and $p$ $+$ Pb collisions with parton cross sections of 3 and 10 mb for different $N_{ch}$ classes \footnote {In principle, the EbyE distribution of $v_n$, P($v_n$), should be obtained by an unfolding method to suppress the nonflow contribution. To our knowledge, the response matrix in the unfolding method cannot be reliably obtained for small colliding systems. Therefore, for consistency, we did not use the unfolding method for both large and small colliding systems.}. Based on P($v_2$, $v_3$), the percentage of the overlapping volume of P($v_2$, $v_3$) between Pb $+$ Pb collisions and $p$ $+$ Pb collisions as a function of the number of final charged particles $N_{ch}$ can be calculated, which is shown in Fig.~\ref{overlap}. Note that the result for 0 mb parton cross section is not shown, since $v_n$ come from non-flow for this case. We can see that the overlapping percentage decreases with parton cross section, which indicates that $v_2$ and $v_3$ between two systems are more different with a larger parton cross section. This is also consistent with the result of test accuracy in Fig.~\ref{ACS}. It suggests that a more pronounced difference in collective flow between the two colliding systems is produced by more parton collisions. On the other hand, the large overlapping volume percentage (over $80\%$) indicates that the P($v_2$, $v_3$) distributions between the two systems are so similar that they are difficult to distinguish. This is also in line with the observation that the PCN can not distinguish the two colliding systems very well in EbyE manner, when we train the PCN with input to be the normalized two-momentum of final particles, as shown in Fig.~\ref{ACS}.

\section{Summary and outlook}
\label{summary} 
 
In summary, we employ the point cloud network to identify the events of $p$ $+$ Pb and peripheral Pb $+$ Pb collisions from a multiphase transport model. We reduce the input information for the PCN and observe the resulting changes in accuracy to verify the specific physical features learned by the PCN. Many different features between the two systems are learned and captured by the PCN, such as pseudorapidity distribution, $p_{\rm T}$ spectra, and anisotropic flow.  In four-dimensional momentum space, the point cloud network can well identify the two different colliding systems. In the transverse momentum plane, the point cloud network can learn the different features of $p_{\rm T}$ spectra that can classify two different colliding systems. After normalizing the transverse momentum of final hadrons, the point cloud network finally distinguishes two different colliding systems according to the feature of collective flow. In this big data and ML approach, by changing the different input types, we confirm that the discrepancy between the two systems is more reflected in the pseudorapidity distribution and the $p_{\rm T}$ spectra than in the anisotropic flow.

Despite these successes, the PCN faces challenges in distinguishing the two systems solely based on event-by-event (EbyE) collective flow, as the EbyE distributions of collective flow parameters $P(v_2, v_3)$ are quite similar between $p$ $+$ Pb and Pb $+$ Pb collisions. However, the PCN could differentiate between ensembles of each system through features related to $v_2$ and $v_3$, and it also revealed the dependence of these discrepancies on $N_{ch}$ and parton cross section. Notably, our findings indicate that the differences in collective flow between $p$ $+$ Pb and Pb $+$ Pb collisions become more pronounced with larger parton scattering cross sections, consistent with the escape mechanism characteristics for collective flow in the transport model~\cite{He:2015hfa,Lin:2015ucn}.

While our PCN has shown some ability to distinguish between $p$ $+$ Pb and peripheral Pb $+$ Pb collisions using event-by-event (EbyE) collective flow analysis, its current effectiveness is somewhat limited. Nevertheless, the application of machine learning has yielded some promising results. This study should be considered a preliminary analysis that lays the foundation for future work involving diverse models and methodologies. However, we emphasize that further research is essential to achieving our ultimate goal, i.e., utilizing machine learning to assist experiments in determining whether small systems also produce the QGP and investigating whether these systems exhibit different characteristics of collective flow compared to large systems. On the other hand, we hope that our PCN study can be applied to classify the origin of atmospheric particle showers, which could improve our understanding of the origins of cosmic rays. Additionally, we also have some prospects for further development of PCN, which holds potential as a valuable tool for various other research interests. For instance, it could be used for searching for the chiral magnetic effect~\cite{Fukushima:2008xe,Zhao:2021yjo} and exploring the nuclear deformation structure~\cite{Jia:2022qgl,Zhao:2022grq} in high-energy heavy-ion collisions.

\begin{acknowledgments}

We thank Dr. Ling-Xiao Wang for helpful discussions. This work is supported in part by the National Natural Science Foundation of China under Grants  No. 12325507, No.12147101, No. 11890714, No. 11835002, No. 11961131011, No. 11421505, No. 12105054, the National Key Research and Development Program of China under Grant No. 2022YFA1604900, the Strategic Priority Research Program of Chinese Academy of Sciences under Grant No. XDB34030000, and the Guangdong Major Project of Basic and Applied Basic Research under Grant No. 2020B0301030008 (S. G., H.-S.W.and G.-L.M.), the CUHK-Shenzhen university development fund under grant No. UDF01003041 and the BMBF funded KISS consortium (05D23RI1) in the ErUM-Data action plan (K. Z.).

\end{acknowledgments}

\bibliography{ref}
\end{document}